\documentclass[%
reprint,
prl,
twocolumn,
 amsmath,amssymb,
 aps,
floatfix,
]{revtex4-1}

\usepackage{graphicx}
\usepackage{dcolumn}
\usepackage{bm}
\usepackage{hyperref}

\usepackage{soul}

\usepackage{graphicx}
\usepackage{times}
\usepackage[normalem]{ulem}
\usepackage{color}

\newcommand{\comment}[1]{}
\renewcommand\sout{\bgroup \color{red} \ULdepth=-.5ex \ULset}
\def\simge{\mathrel{\rlap{\raise 0.511ex
     \hbox{$>$}}{\lower 0.511ex \hbox{$\sim$}}}}
\def\simle{\mathrel{\rlap{\raise 0.511ex
      \hbox{$<$}}{\lower 0.511ex \hbox{$\sim$}}}}

\graphicspath{{./}{figures/}}

\begin{document}

\title{The Properties of a Black Hole-Neutron Star Merger Candidate}


\author{James M. Lattimer$^1$}
\affiliation{$^1$ Department of Physics and Astronomy, Stony Brook University, Stony Brook, NY 11794-3800, USA}
 
\begin{abstract}
The LIGO/Virgo Consortium (LVC) released a preliminary announcement of
a candidate gravitational wave signal, S190426c, that could have
arisen from a black hole-neutron star merger.  As the first such
candidate system, it's properties such as masses and spin are of great
interest.  Although LVC policy prohibits disclosure of these properties in
preliminary announcements, LVC does release the estimated probabilities that
this system is in specific categories, such as binary neutron star,
binary black hole and black hole-neutron star.  LVC also releases information
concerning relative signal strength, distance, and the probability
that ejected mass or a remnant disc survived the merger.  In the case
of events with a finite probability of being in more than one
category, such as is likely to occur with a black hole-neutron star
merger, it is shown how to estimate the masses of the components
and the spin of the black hole.  This technique is applied to the
source S190426c.
\end{abstract}

\maketitle


\section{Introduction}
On April 26, 2019 The LIGO/Virgo Consortium (LVC) observed
gravitational waves from a possible compact object coalescence,
S190426c~\cite{LVC19a, LVC19b} that could be the first observed black
hole-neutron star (BHNS) system.  GCN circular 24237~\cite{LVC19a}
reported a false alarm probability of 1.9e-08 Hz or once every 1 year,
7 months, which indicates a moderately low signal-to-noise (S/N)
ratio, supported by the 14\% estimated probability of being a
terrestrial anomaly.  Initially, the system was assigned a probability
$p_{\rm BNS}=57\%$ of being a binary neutron star (BNS), $p_{\rm
  BHNS}=15\%$ of being a BHNS system, $p_{\rm gap}=28\%$ of being a
MassGap system (henceforth called 'gap'), and $p_{\rm BBH}<1\%$ of
being a binary black hole system (BBH), assuming it is cosmic in
origin.  These classifications are based on the convention that
component masses less than $3M_\odot$ are neutron stars and those
greater than $5M_\odot$ are black holes.  Systems with one or two
components in the gap between $3M_\odot$ and $5M_\odot$ are classified
as gap.  In addition, it was reported that the probabilities that mass
existed outside of the compact remnant at least briefly after the
coalescence, HasRemnant (henceforth $p_d$), and that at least one
component was a neutron star, HasNS (henceforth $p_{\rm NS}$), were
both $>99$\%.  The source classification probabilities, but not the
source distance, false alarm rate, terrestrial anomaly probability,
and $p_{\rm NS}$, were later revised in GCN circular
24144~\cite{LVC19b} to be $p_{\rm BNS}=15\%$, $p_{\rm BHNS}=60\%$,
$p_{\rm gap}=25\%$ and $p_{\rm BBH}<1\%$, assuming the source is
cosmic in origin.  In addition, $p_d$ was revised to 72\%, indicating
that mass ejection is considered likely.  As originally proposed in
Ref.~\cite{Lattimer74}, some BHNS mergers could result in mass
ejection leading to the synthesis of r-process
nuclei~\cite{Lattimer76,Lattimer77}, and an optical
signature~\cite{Li98,Metzger10}.  It is therefore of great interest to
understand more about this system, including its component masses.

It should be noted that the neutron star maximum mass is undoubtedly
less than about $3M_\odot$ due to causality, so that gap objects are
presumably black holes.  Therefore, gap systems are, in conventional
notation, either BHNS or BBH systems.  Since the system has greater
than 99\% chance of containing one neutron star, the reported
probabilities indicate that there is actually an 85\% chance of this
being a BHNS system.

In principle, the system's chirp mass ${\cal
  M}=(M_1M_2)^{3/5}/(M_1+M_2)^{1/5}$ ought to have been determined to
relatively high precision even in the case of a BHNS merger with a
relatively low S/N ratio.  With a S/N ratio approximately 3,
extrapolating the results of Ref.~\cite{Lackey12} indicates
$\Delta{\cal M}/{\cal M}\sim0.003$.  However, in all likelihood, the
binary mass ratio $q=M_1/M_2\ge1$ and component spins were determined
relatively poorly; a similar extrapolation indicates
$\Delta\eta/\eta\sim0.5$ where $\eta=q/(1+q)^2$ is the symmetric mass
ratio.  Unfortunately, current collaboration policy forbids the
release of ${\cal M}, q$ and spin estimates or their uncertainties.
Therefore, it is not possible 
to learn the likely source masses or spins until publication of a
peer-reviewed article, which may not occur for several months
following a detection.  Fortunately, it possible to estimate the
system masses and black hole spin from the information released in the
GCN circulars with a few straightforward assumptions.

\begin{figure}[t]
\vspace*{-.75cm}
\hspace*{-1.05in}\includegraphics[width=.7\textwidth,angle=180]{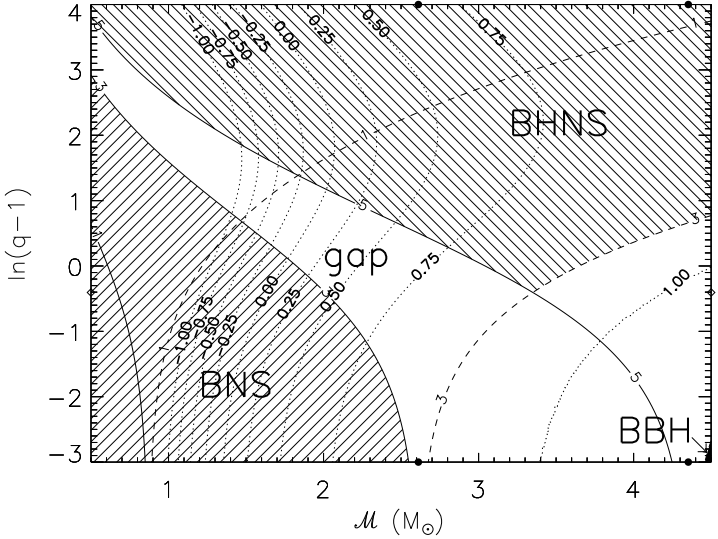}
\vspace*{-.6in}
\caption{Compact binary classifications in the ${\cal M}-\bar q$ plane.  BNS and BHNS regions are hatched, the gap region is unshaded, and the BBH region is solid.  Labelled contours of $M_1 (M_2)$ in $M_\odot$ are shown as solid (dashed) curves.  Dotted contours are the boundaries $M_d=0$ labelled for various values of black hole spin parameter $\chi$. Filled points show the maximum (minimum) ${\cal M}$ values for BNS (BBH) systems.  Diamonds show the minimum $\bar q$ value for BHNS systems.\label{fig:class}}
\end{figure}

\vspace*{-.2cm}\section{Method}
The case of BHNS mergers is especially interesting, because at least three classification categories likely then have finite
probabilities which provides additional information compared
to BNS or BBH events.  It is convenient to examine this problem, not
in $M_1-M_2$ space, but in ${\cal M}-q$ or ${\cal M}-\eta$ space,
because ${\cal M}$ presumably has a very small uncertainty.
However, due to their expected large uncertainties, $q$ or $\eta$ are
not suitable variables given their ranges $q\in[1,\infty]$ and
$\eta\in[0,1/4]$ for which Gaussian probability distribution with
any finite uncertainty $\sigma_q$ or $\sigma_\eta$ will extend into
the non-physical regions $q<1$ or $\eta>1/4$.  It is convenient to
employ the alternate variable $\bar q=\ln(q-1)$ which has the range
$\bar q\in[-\infty,\infty]$. Note that $\bar q=[-1,0,1]$ corresponds
to $q=[1.37,2,3.72]$.  We will refer to the uncertainty in $\bar q$ by
$\sigma_q$ for notational simplicity. Fig. \ref{fig:class} shows
classifications in ${\cal M}-\bar q$ space, together with some
$M_1$ and $M_2$ contours.

Given measurements ${\cal M}_0\pm\sigma_{\cal M}$ and $\bar q_0\pm\sigma_q$, we
take the probability density of an event having ${\cal M}$ and
$\bar q$ to be
\begin{equation}
\vspace*{.25cm}
{d^2p\over d{\cal M}d\bar q}=A\exp\left[-{({\cal M}-{\cal M}_0)^2\over2\sigma_{\cal M}^2}-{(\bar q-\bar q_0)^2\over2\sigma_{q}^2}\right]
\end{equation}
where $A=(2\pi\sqrt{\sigma_{\cal M}\sigma_{q}})^{-1}$.  It is assumed
that the uncertainties in ${\cal M}$ and $\bar q$ are uncorrelated.
One expects that $\sigma_{\cal M}$ will be very small, possibly of order
a few hundreths of a solar mass or less, while $\sigma_{q}$ will be
large, of order unity or larger.  Therefore, results should be insensitive to $\sigma_{\cal M}$.

\begin{figure}[h]
\vspace*{-.15cm}
\hspace*{-0.7in}\includegraphics[width=.61\textwidth,angle=180]{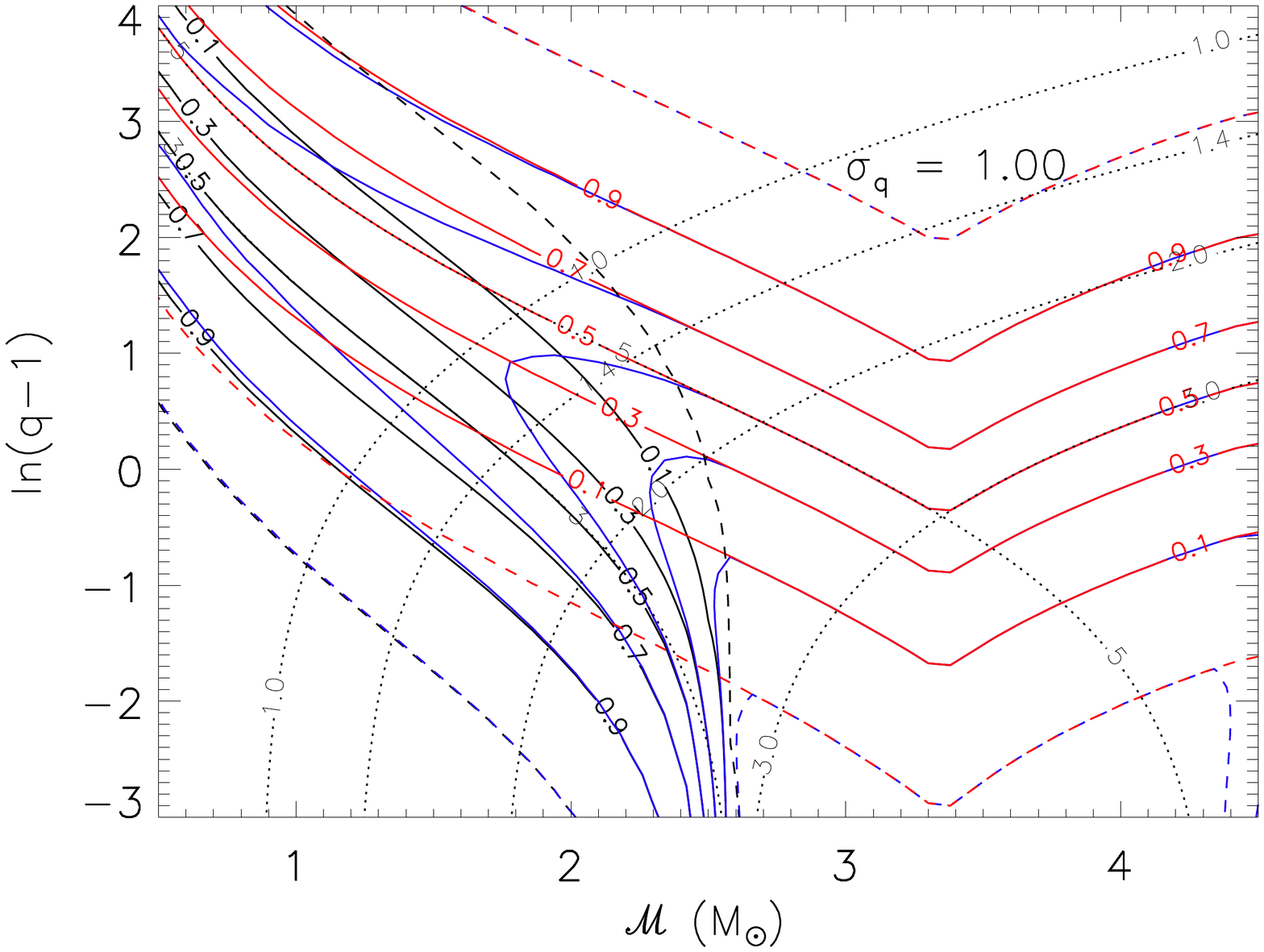}
\hspace*{-1.in}\includegraphics[width=.61\textwidth,angle=180]{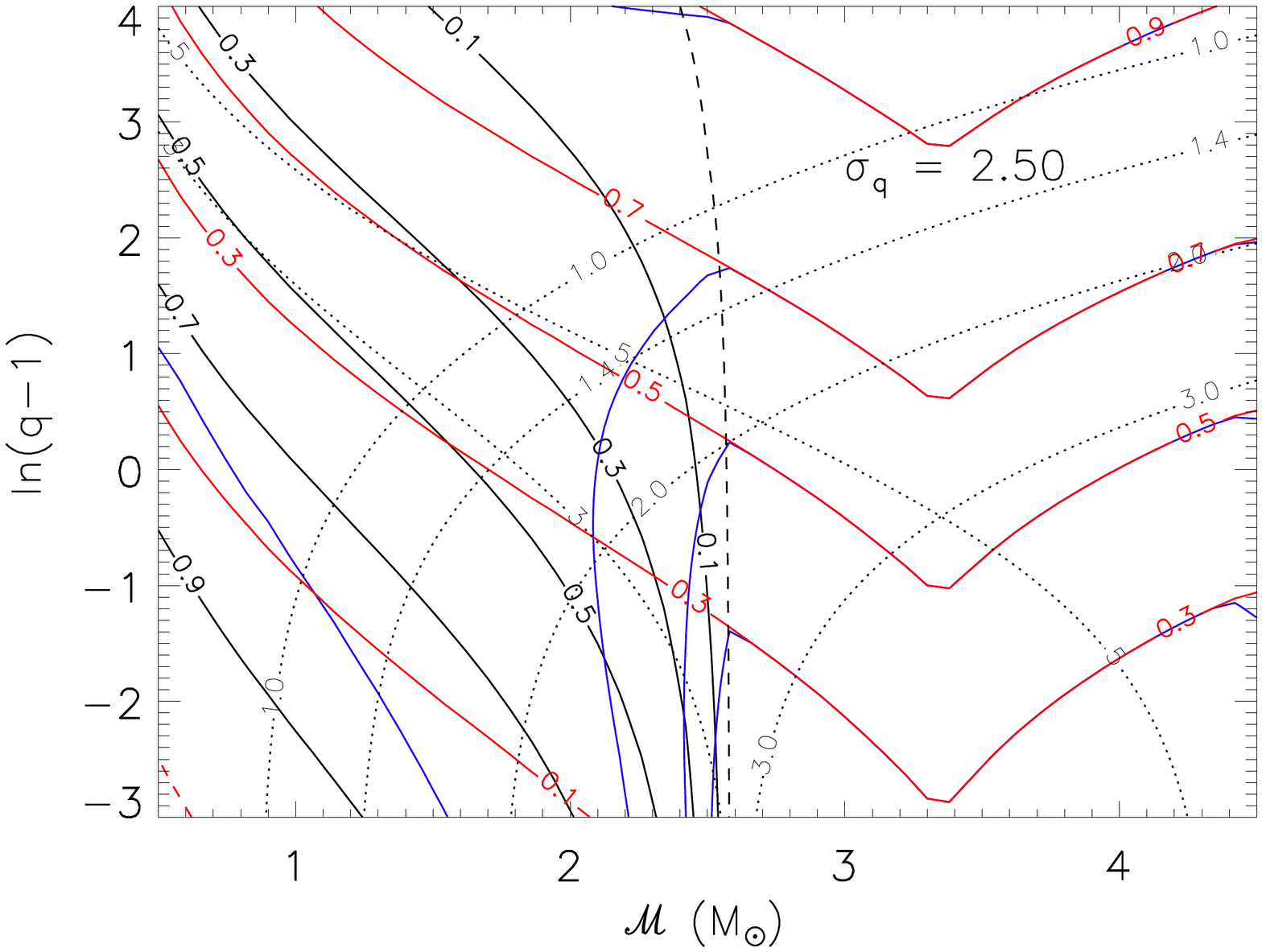}
%
\vspace*{-.15cm}
\caption{Classification probability contours for two values of
  $\sigma_q$.  In all cases, $\sigma_{\cal M}=0.01{\cal M}$.  Solid
  red (black) contours show probabilties for BHNS (BNS) systems; blue
  solid contours show gap probabilities.  Only BHNS and BNS contours
  are labelled.  Dashed contours indicate probabilities of 0.01 or
  0.99 for each class.  Dotted curves are contours of $M_1
  (3M_\odot,5M_\odot)$ and $M_2
  (M_\odot,1.4M_\odot,2M_\odot,3M_\odot)$.  \label{fig:prob}}
\end{figure}

For given values of ${\cal M}_0$ and
$\bar q_0$, the integration of the probability density over all values
of ${\cal M}$ and $\bar q$ corresponding to the BNS,  gap, BHNS and BBH
regions yields the probabilities $p_{\rm BNS}, p_{\rm gap}, p_{\rm BHNS}$ and $p_{\rm BBH}$, 
respectfully, which will add to unity.  Each of
these probabilities are therefore functions of four variables $[{\cal
    M},\sigma_{\cal M},\bar q,\sigma_{q}]$.  A particular choice of
$\sigma_{\cal M}$ and $\sigma_{q}$ then allows probability contours to
be drawn in ${\cal M}-{\bar q}$ space.  However, because $\sigma_{\cal M}$ is
relatively small, these contours are insensitive to $\sigma_{\cal M}$
and depend primarily on $\sigma_{q}$.  Fig. \ref{fig:prob} shows these
probability contours for $\sigma_{\cal M}=0.01{\cal M}$ for the 
cases $\sigma_{q}=[1.0,2.5]$.

Realistically, neutron stars have a minimum mass around
$1M_\odot$, but the LVC algorithm doesn't consider this minimum
when assigning classification probabilities.  Therefore, there is no
lower bound to ${\cal M}$ for BNS although there is an upper bound.
Because $\sigma_{\cal M}$ is small, 
$p_{\rm BNS}>0.01$ requires ${\cal M}/M_\odot<3/2^{1/5}=2.612$.
Similarly, $p_{\rm BBH}>0.01$ requires ${\cal
  M}/M_\odot>5/2^{1/5}=4.353$. There are no ${\cal M}$ bounds for gap
or BHNS events with finite probabilities.  But a minimum value $q>5/3$
is required for $p_{\rm BHNS}\simge0.01$. \phantom{xxx}\hspace*{1.6in}

\onecolumngrid

\twocolumngrid
For systems with a
finite chance of being a BNS binary, $p_{\rm NS}$ is always greater than 0.99 and therefore provides no additional information to constrain ${\cal M}$ and $\bar q$. For small $\bar q$, the $p_{\rm NS}=0.99$
contour lies close to the limiting curve for the smaller component to
be a neutron star, i.e., $M_2=3M_\odot$ (see Fig. \ref{fig:s190426c}).
But for larger values of $q$, especially for large $\sigma_q$, this
contour becomes more vertical than the $M_2=3M_\odot$ contour.  For
$\bar q\simge0$, the $p_{\rm NS}=0.99$ contour always lies to the right of that for
$p_{BNS}=0.01$. 

Another quantity provided~\cite{LVC19b} is the
probability, $p_{d}$, of having disrupted material outside the merged
object.  This, together with the determination of ${\cal M}$ and $\bar q$
with the above method, allows an estimate of the black hole spin by
using an analytic model for the mass $M_d$ of disrupted
material~\cite{Foucart12,Foucart18}.  The same model, which has three free
parameters and was fit to relativistic hydrodynamical
results, was used by LVC to determine $p_d$ from their inferred
values of ${\cal M}$ and $q$~\cite{Farr19}.  It approximates the
combined mass, $M_d$, of the accretion disk, the tidal tail, and the
potential ejecta, remaining outside the black hole a few milliseconds
after a BHNS merger, by
\begin{equation}
M_d\simeq M_{\rm b,NS}\left[{\alpha^\prime\over\eta^{1/3}}\left(1-2\beta\right)-{\beta^{\prime}\over\eta}\hat R_{\rm ISCO}\beta+\gamma^\prime\right],
\label{eq:md}\end{equation}
where $\beta=GM_{\rm NS}/(R_{\rm NS}c^2)$ is the neutron star compactness,
$R_{\rm NS}$ is the neutron star radius, $M_{\rm b,NS}$
is the baryon mass of the neutron star, and $\hat R_{\rm ISCO}=R_{\rm ISCO}c^2/(GM_{\rm BH})$.  $R_{\rm ISCO}$ is the radius
of the innermost stable circular orbit, from which the black hole spin
parameter $\chi$ may be found:
\begin{equation}
\chi={\sqrt{\hat R_{\rm ISCO}}\over3}\left({4}-\sqrt{{3\hat R_{\rm ISCO}}-{2}}\right).
\label{eq:chi}\end{equation}
The model is claimed to be accurate to within a few percent of the
mass of the neutron star, $M_{\rm NS}$.  The model constants are
$\alpha^\prime\simeq0.406$, $\beta^\prime\simeq0.139$, and
$\gamma^\prime=0.255$.  LVC have assumed $R_{\rm NS}=15$
km~\cite{Farr19}.  The boundary of the ${\cal M}-\bar q$ plane
permitting non-zero $M_d$, is then found from
\begin{equation}
\hat R_{\rm ISCO}\simeq(\beta^\prime\beta)^{-1}\left(\alpha^\prime\eta^{2/3}(1-2\beta)+\gamma^\prime\eta\right),
\label{eq:risco}\end{equation} 
which contains only two model parameters.
  The boundaries for various values of $\chi$ are shown
in Fig. \ref{fig:class}; the regions to the left of each boundary are where $M_d>0$.
It should be noted that values of $\chi<-1$ and $\chi>1$ are possible solutions of Equations (\ref{eq:chi}) and (\ref{eq:risco}).  These admittedly unphysical regions are unlikely to be populated by BHNS mergers, as judged by numerical relativity simulations.

The probability $p_{d}$ for a given value of $\chi$ can be determined by integrating $d^2p/d{\cal
  M}d{\bar q}$ over the region in ${\cal M}-\bar q$ space permitting non-zero $M_d$.  When
$p_{d}$ is provided, the likely black hole spin is
found by observing which $\chi$ contours corresponding to this value of
$p_{d}$ pass through the favored $[{\cal M},\bar q]$ region determined
by satisfying the $p_{\rm BNS}, p_{\rm gap}$ and $p_{\rm BHNS}$ conditions.  As is the case for ${\cal M}$ and $\bar q$
themselves, the inferred value of $\chi$ is insensitive to $\sigma_{\cal M}$ and depends mostly on $\sigma_{\bar q}$.  




%

\section{Application to S190426c}

\begin{figure}[h]
\vspace*{-.3in}
\hspace*{-0.7in}\includegraphics[width=.61\textwidth,angle=180]{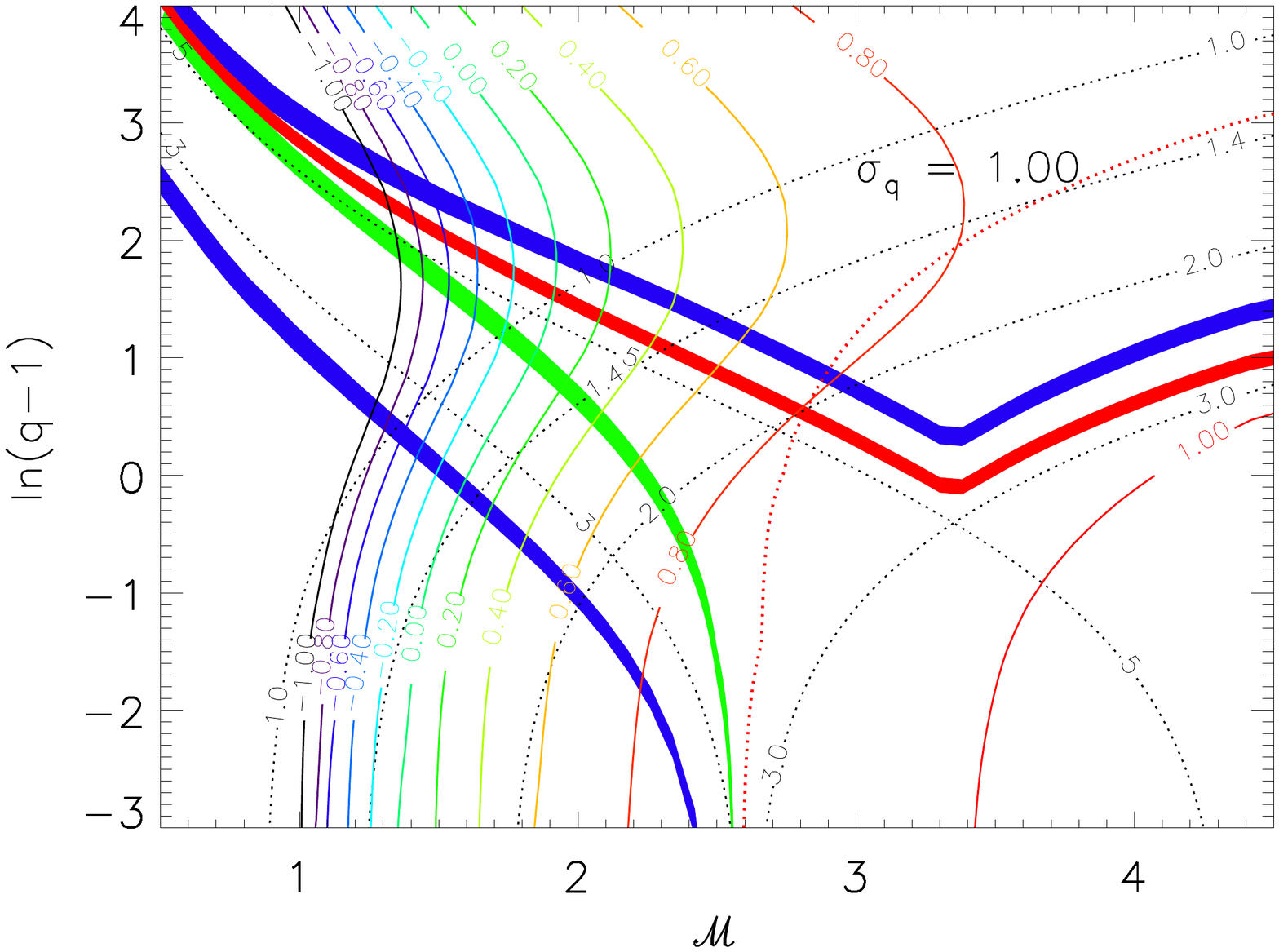}
\hspace*{-1.in}\includegraphics[width=.61\textwidth,angle=180]{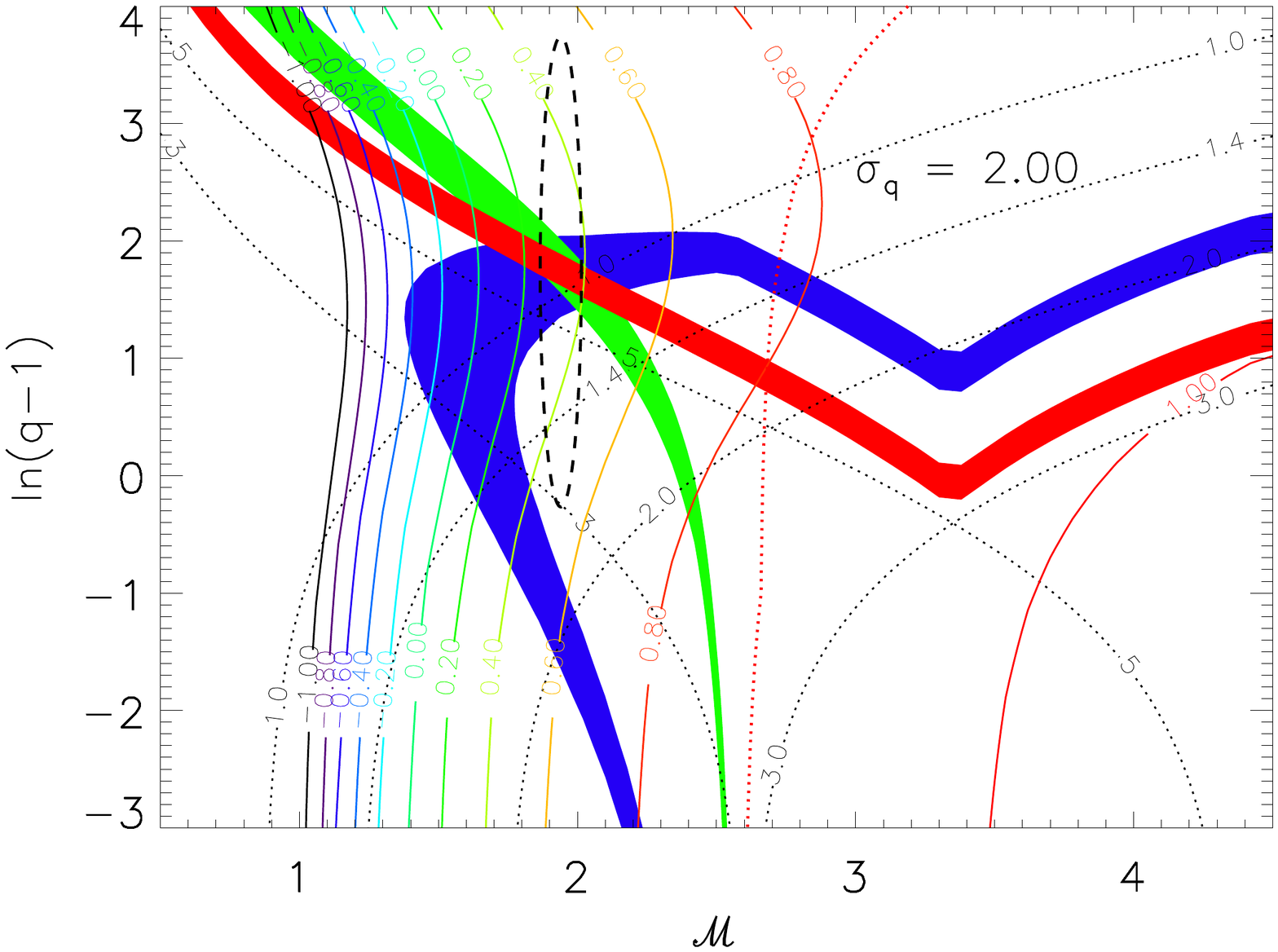}

\vspace*{-.75in}
\hspace*{-0.7in}\includegraphics[width=.61\textwidth,angle=180]{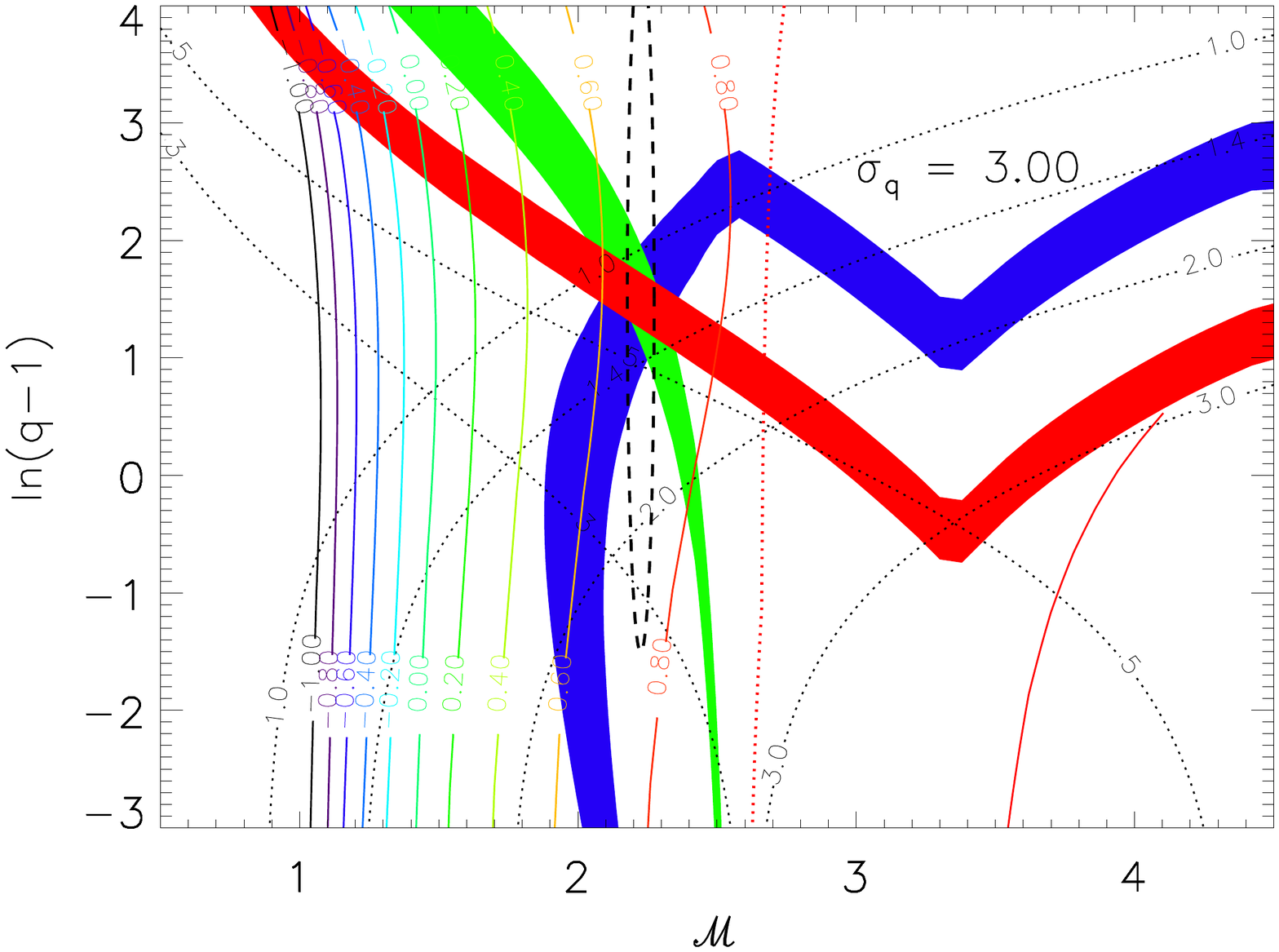}
\hspace*{-1.in}\includegraphics[width=.61\textwidth,angle=180]{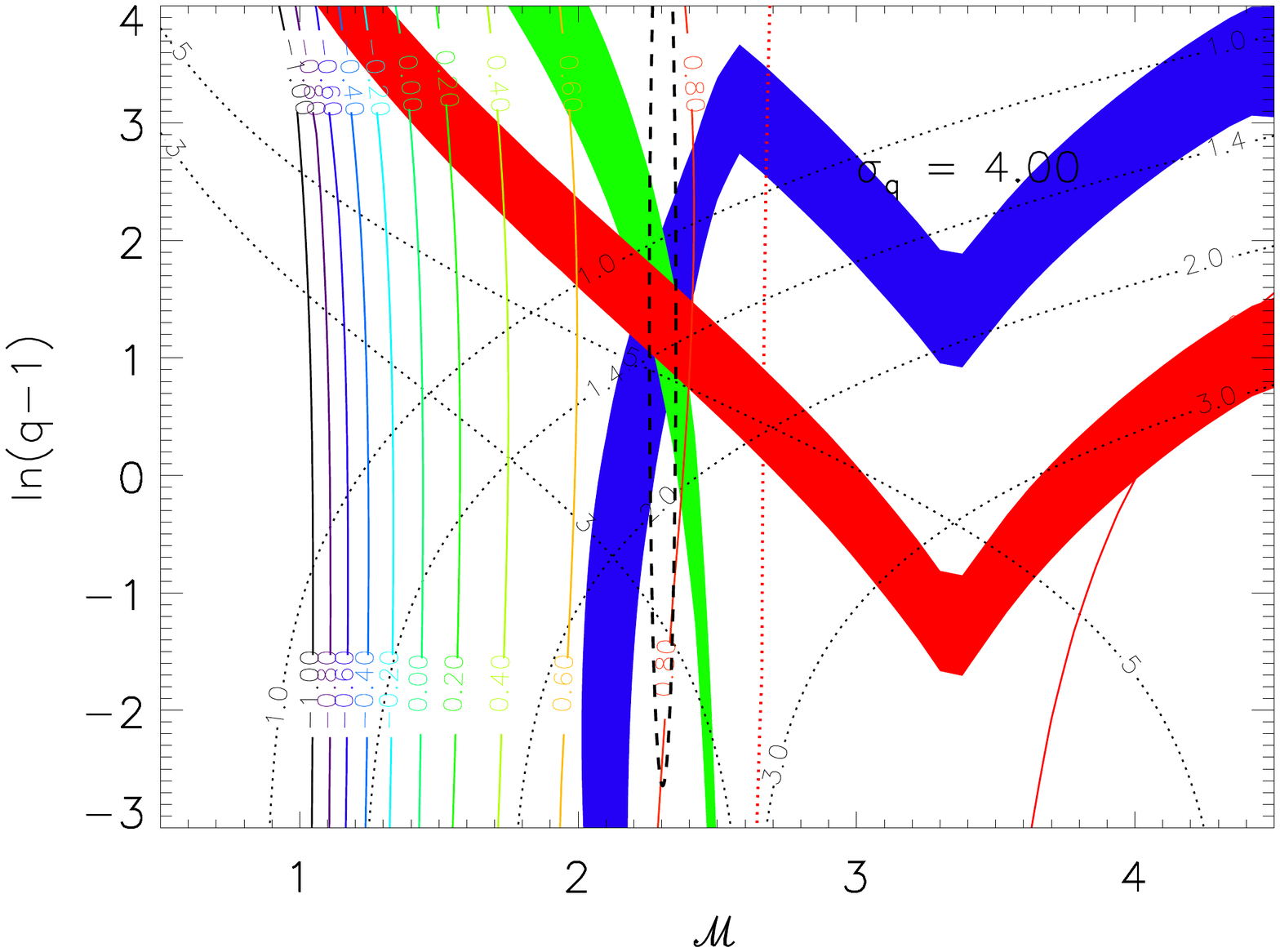}
%
\vspace*{-.3in}
\caption{Probability regions matching those reported by LVC for the possible BHNS event S190426c for four assumed values of $\sigma_q$.  The green, blue and red regions show where $p_{\rm NS}=0.150\pm0.025$, $p_{\rm gap}=0.250\pm0.025$, and $p_{\rm BHNS}=0.600\pm0.025$, respectively.  Labelled solid contours show the black hole spin parameters $\chi$ where $p_d=0.72$.  The red dotted contour is $p_{\rm NS}=0.01$.  The dashed ellipsoid represents the inferred $1\sigma$ confidence elllipse consistent with all three probabilities.  
\label{fig:s190426c}}
\end{figure}

The event S190426c~\cite{LVC19b} was reported to have $p_{\rm
  BNS}=3/20$, $p_{\rm gap}=5/20$, $p_{\rm BHNS}=12/20$, $p_d=0.72$, and
$p_{\rm NS}>0.99$ (which is not
useful, as previously described).  We assume that the classification probabilities
are uncertain by $\pm0.5/20$.  One can then identify the most
likely values of ${\cal M}$ and $\bar q$ by plotting regions defined
by $p_{\rm BNS}=0.15\pm0.025$, $p_{\rm gap}=0.25\pm0.025$ and
$p_{\rm BHNS}=0.60\pm0.025$ and identifying any overlap regions
(Fig. \ref{fig:s190426c}). It is assumed that $\sigma_{\cal
  M}=0.01{\cal M}$ based on likely expectations for even small S/N
events, but in any case these contours are very insensitive to this
parameter as long as $\sigma_{\cal M}<0.1{\cal M}$.  This
insensitivity is not the case for $\sigma_q$, however.  Results for
selected choices of $\sigma_q$ are shown in
Fig. \ref{fig:s190426c}.  For $\sigma_q\simge1.25$, consistent
solutions become possible.

\begin{figure}[h]
\vspace*{-.75cm}
\hspace*{-0.75in}\includegraphics[width=.65\textwidth,angle=180]{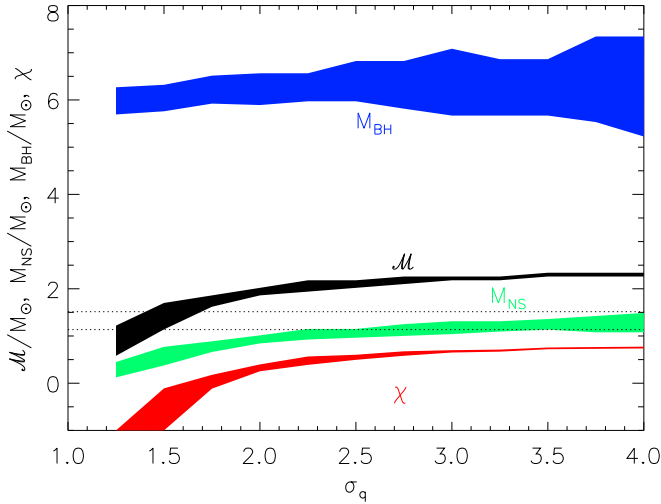}
\vspace*{-.5in}
\caption{Inferred ${\cal M}$, $M_{\rm NS}$, $M_{\rm BH}$ and $\chi$
  for S190426c as functions of $\sigma_q$.
The dotted lines indicate the 95\% confidence interval determined from the observed binary neutron star mass distribution (see text).
\label{fig:par}}
\end{figure}

When a consistent solution is possible, a region in ${\cal M}-\bar q$ space is
outlined.  The overall uncertainties are established
by combining the size of this region with the assumed values of
$\sigma_{\cal M}$ and $\sigma_q$.  The implied ${\cal M}-\bar q$ confidence ellipses
 are shown in Fig. \ref{fig:s190426c}.  While the individual masses have large uncertainties
because $\sigma_q$ is large, the inferred centroids of the black hole
mass for those cases where a consistent solution exists is relatively
independent of the actual value of $\sigma_q$, and is around
$6M_\odot$ (Fig. \ref{fig:par}).  The neutron star mass, on the
other hand, increases progressively up to $\sim1.4M_\odot$
as $\sigma_q$ is increased.

These results suggest the existence in S190426c of a relatively
low-mass neutron star.  For the smallest values of
$\sigma_q\simeq1.25$ for which a consistent solution is possible, the
inferred mass, about $0.25M_\odot$, is less than the minimum possible
neutron star mass, about $1.1M_\odot$, that can be made in core-collapse
supernova events~\cite{Strobel99,Suwa18}.  The lowest well-measured neutron star mass, the
companion to PSR J0453+1559, is $1.174\pm0.004M_\odot$~\cite{Martinez15}.
The observed BNS population currently consists of 9 systems with
well-measured individual masses and 7 systems with well-measured total
masses $M_T$.  Assuming these are sampled from a Gaussian
distribution, Ref.~\cite{Ozel16} obtained a mean mass
$1.33\pm0.09M_\odot$ from fitting 7 of the 9 systems with
well-measured individual masses.  Including all 16 systems one finds
the nearly identical result, $1.325\pm0.095M_\odot$. The systems for
which only the total mass is known can be treated assuming the lower
(higher) mass star cannot have a mass greater (less) than $M_T/2$, and
that the minimum neutron star mass is $M_{min}=1.1M_\odot$, as argued above.
Assuming these systems are more likely to be symmetric than highly
asymmetric, as observed for other BNS systems, it seems justified to
assume that the lower mass component has mass $M_2$ with a probability
proportional to $M_2-M_{min}$ for $M_{min}<M_2<M_T/2$, and the
higher mass component has mass $M_1$ with a probability proportional
to $M_T-M_{min}-M_1$ for $M_T/2<M_1<M_T-M_{min}$.  The 95\%
confidence bounds for this distribution,
$[1.135M_\odot,1.515M_\odot]$, estimated using twice the standard
deviation $0.095M_\odot$, are indicated in Fig. \ref{fig:s190426c}.
Values of $\sigma_q\simge2$ then predict a neutron star mass
consistent with the observed distribution.

From the value $p_d=0.72$, the inferred black hole spin $\chi$
values, with uncertainties determined from the size of the overlap
region in ${\cal M}-\bar q$ space, are shown as a function of
$\sigma_q$ in Fig. \ref{fig:par}, along with inferred ranges of ${\cal
  M}$, $M_{\rm NS}$, and $M_{\rm BH}$.  Note that for
$\sigma_q\simle1.5$, the minimum $\chi$ may fall below -1.  As this is
unphysical, it strongly supports larger values of $\sigma_q$, which is
consistent with the results implied by realistic values of $M_{\rm NS}$.

\phantom{unecessary text for formatting}
\onecolumngrid

\twocolumngrid

\section{Discussion}
 It is interesting that, irrespective of assumptons concerning
 $\sigma_q$, we predict that $M_{\rm BH}\simeq6M_\odot$, and, in the
 case of moderate to large values of $\sigma_q$, we find that $M_{\rm
   NS}$ converges to the range expected from observed binary neutron
 star masses.  LVC initially reported that the probability that this
 event was terrestrial, i.e., that it is spurious, to be 14\%, and
 have not updated that estimate.  The fact that we find physically
 realistic values for $M_{\rm NS}$ seems to lend a degree of credulity
 to the real nature of this event and supports moderate to large
 values of $\sigma_q$, with an inferred mass ratio $q\sim4$.  On the
 other hand, the inferred $\chi$ values for large $\sigma_q\simge3$
 approach 0.75.  Such rapidly spinning black holes are
 inconsistent with smaller values inferred from BBH
 mergers~\cite{Bavera19}.  The mean value of the effective binary spin
 $\chi_{eff}=(M_1\chi_1+M_2\chi_2)/(M_1+M_2)$ for the first 10 BBH
 observed by LVC was $0.046\pm0.052$. Smaller values of
 $\chi\sim0$ would favor $1.75\simle\sigma_q\simle2.25$,
 suggesting that $M_{\rm NS}$ is near the lower end of the
 range of observed masses.

LVC overestimates the probability of a surviving remnant.  As previously noted, LVC assumes a value $R_{\rm NS}=15$ km
to estimate $M_d$ from Equation (\ref{eq:md}).  This radius value is
unrealistically large, given evidence from nuclear experiment and
theory~\cite{Lattimer13}, as well as from the tidal deformations inferred
from GW170817~\cite{De18,Abbott18}.  A more realistic value $R_{\rm
  NS}\simeq12$ km makes the probability of a surviving disc much less likely since the resulting $\chi$ contours in Figs.
\ref{fig:class} and \ref{fig:s190426c} would be shifted considerably
to the left.  Changing $R_{\rm NS}$ to 12 km would change the condition needed for $M_d>0$ from $\chi>0.25$ for our suggested solution ${\cal M}\simeq2.3M_\odot$ and $\bar q\simeq1.5$ (see Fig. \ref{fig:class}) to $\chi>0.60$.  For $\sigma_q$ in the range 1.75 - 2.25, one then finds the condition $p_d\simge0.0$ requires $\chi\simge0.5$, considerably larger than our favored value near zero.   In other words, the use of
a more realistic neutron star radius together with our inferred ${\cal
  M}$ and $\bar q$ values implies that the formation of a remnant disc
is much less likely than announced.  It is thought that if a disc cannot form, tidal or wind ejection
of matter is also unlikely, and the resulting synthesis of radioactive
heavy nuclei and subsequent optical emission is not possible.
This would be consistent with the apparent failure to observe an
electromagnetic counterpart.

\vspace*{0.25in}
\section{Acknowledgements}
I thank Will Farr for helpful discussions.  This work was supported by DOE Award DE-FG02-87ER40317.

\bibliography{bhnsprob}

\end{document}